\documentclass[aps,prd,preprint,superscriptaddress,tightenlines,nofootinbib,showpacs]{revtex4}
\usepackage{epsfig}
\usepackage{dcolumn}% Align table columns on decimal point
\usepackage{bm}% bold math

\def\lamctolam4pi{\Lambda_c^+\to \Lambda\pi^+\pi^+\pi^-\pi^0}
\def\lamctolamompi{\Lambda_c^+\to \Lambda\omega\pi^+}
\def\lamctolametapi{\Lambda_c^+\to \Lambda\eta\pi^+}
\def\lamctopkpi{\Lambda_c^+\to p K^-\pi^+}
\def\lamctolampi{\Lambda_c^+\to \Lambda\pi^+}

\begin{document}
\preprint{\tighten\vbox{\hbox{\hfil CLNS {02/1800}}
                        \hbox{\hfil CLEO {02-13}}
}}

%\draft % needed for PACS to appear
%\tighten

\title{
First Observation of the Exclusive Decays $\lamctolam4pi$ and $\lamctolamompi$
}
\author{D.~Cronin-Hennessy}
\author{A.L.~Lyon}
\author{C.~S.~Park}
\author{W.~Park}
\author{J.~B.~Thayer}
\author{E.~H.~Thorndike}
\affiliation{University of Rochester, Rochester, New York 14627}
\author{T.~E.~Coan}
\author{Y.~S.~Gao}
\author{F.~Liu}
\author{Y.~Maravin}
\author{R.~Stroynowski}
\affiliation{Southern Methodist University, Dallas, Texas 75275}
% ...
%\author{(CLEO Collaboration)} %FOR PRD
\collaboration{CLEO Collaboration} %FOR PRL,CLNS
\noaffiliation

% You will want to hard code the date once you are ready to submit your paper!
\date{October 18, 2002}

\begin{abstract} 
Using data collected by the CLEO III detector at CESR, we report on
the first observation of the decays $\lamctolam4pi$ and $\lamctolamompi$. The branching fractions are measured relative 
to the $pK^-\pi^+$ mode; we find ${\cal B}( \lamctolam4pi ) / {\cal B}( \lamctopkpi) = 0.36\pm$0.09(stat.)$\pm$0.09(syst.)
and ${\cal B}( \lamctolamompi )/{\cal B}( \lamctopkpi )$ = 0.24$\pm$0.06(stat.)$\pm$0.06(syst.). 
We also observe a small excess of events in $\lamctolametapi$, $\eta\to\pi^+\pi^-\pi^0$ from which we
find ${\cal B}( \lamctolametapi ) / {\cal B}( \lamctopkpi)$ = 0.41$\pm$0.17$\pm$0.10 with a corresponding 
limit of $<$0.65\% at 90\% confidence level. This rate is consistent with previous measurements of 
$\lamctolametapi$, $\eta\to\gamma\gamma$. We find that the combined branching fractions for $\lamctolamompi$ and 
$\lamctolametapi$ comprise (87$\pm$13)\% of the total $\lamctolam4pi$ rate. We set an upper limit on 
${\cal B}( \lamctolam4pi ) / {\cal B}( \lamctopkpi)$ for non-$\omega$ and non-$\eta$ modes of 0.13
at 90\% confidence level.
\end{abstract}

\maketitle
%\tighten

\setcounter{footnote}{0}

\newpage
%\section{Introduction}\label{sec:Introduction}
   Over the last decade large data samples have been used to measure charm
meson and charm baryon branching fractions to exclusive final states. 
Exclusive final states for $\Lambda_c^+\to\Lambda (n)\pi$ decays, with n=1,2,3 have been measured~\cite{lampi_modes}, but modes with four (or more) pions have yet to be observed. The large rates for $D^+\to \overline{K^0}\pi^+\pi^+\pi^-\pi^0$ 
($5.4^{+3.0}_{-1.4})\%$)~\cite{PDG2000} and $D^0\to K^-\pi^+\pi^+\pi^-\pi^0$ (($4.0\pm 0.4)\%$)~\cite{PDG2000} suggest non-negligible rates for $\lamctolam4pi$. In this paper, we report the first measurement of the branching fractions for $\lamctolam4pi$ and $\lamctolamompi$, $\omega\to\pi^+\pi^-\pi^0$. There is also some indication that a portion
of the $\Lambda\pi^+\pi^+\pi^-\pi^0$ occurs through $\lamctolametapi$, $\eta\to\pi^+\pi^-\pi^0$. We find that
the resonant $\omega$ and $\eta$ sub-modes nearly account for most (if not all) of the $\lamctolam4pi$ rate.
Throughout this paper, charged conjugate states are implied unless otherwise specified.

	The data for this analysis were collected using the CLEO III detector located at the Cornell Electron Storage Ring (CESR). The data sample consists of 2.7 fb$^{-1}$ on or near the $\Upsilon$(4S), 1.4 fb$^{-1}$ on or near the $\Upsilon$(3S), and 0.02 fb$^{-1}$ on or near the $\Upsilon$(1S). The CLEO III detector is an upgrade of the CLEO II/II.V detector~\cite{CLEOdetector} which includes a four-layer double-sided silicon vertex detector, a tracking drift chamber~\cite{dr3} consisting of 16 inner axial layers and 31 outer small-angle stereo layers, and
a ring-imaging Cerenkov detector (RICH)~\cite{rich}. The RICH is located outside the tracking chambers and uses LiF radiators to produce Cherenkov photons from incident charged tracks. The Cerenkov cone expands in a 20-cm thick ${\rm N_2}$-filled expansion volume and are then photo-converted in a gas mixture of Methane and TEA and detected by a multi-wire chamber with cathode pad readout. The electromagnetic calorimeter consists of 7800 CsI crystals which are used to measure the position and energy of photons and to aid in lepton identification. These detector components are immersed in a 1.5 Tesla solenoidal magnetic field that is uniform to within 0.2\% over the tracking volume. Outside the solenoid there is a muon system covering 85\% of the solid angle.

	Signal candidates are formed using charged tracks and photons. We only use charged tracks which have a high quality track fit and have $| \cos\theta | <0.925$, where $\theta$ is the polar angle of the track with respect to the $e^-$ direction. Charged particles (except those used in forming $\Lambda$ candidates)
are required to have a distance of closest approach (to the interaction point) in the $r-\phi$ plane of 5 mm and 5 cm along the $\hat{z}$ axis (the direction of the $e^-$ beam, apart from the small crossing angle of the $e^+e^-$ beams). Photons used in forming $\pi^0$ candidates must be isolated from charged tracks and have a shower shape that is consistent with the shape of an electromagnetic shower. The isolation criterion requires that no charged track points to any of the crystals which are used to make up the electromagnetic shower.

   Particle identification is performed using information on the
specific ionization ($dE/dx$) measured in the drift chamber and particle likelihoods obtained from the RICH. For $dE/dx$, we quantify the
consistency of the measured $dE/dx$ with $p$, $K$ and $\pi$ hypotheses by defining a significance, $S_i=( (dE/dx)^{obs}_i-\langle dE/dx\rangle_i )/\sigma_i(dE/dx)$, where $(dE/dx)^{obs}_i$ and $\langle dE/dx \rangle_i$ are the measured ionization (corrected for path length) and expected (momentum-dependent) mean value of $dE/dx$ for particle type $i$ ($i=p, K, \pi$) respectively, and $\sigma_i(dE/dx)$ is the resolution on the measured $dE/dx$.  For the RICH, we use the measured track momentum and a given particle hypothesis to search for Cerenkov photons whose measured positions are consistent (within 3 standard deviations) with expected positions. For each particle hypothesis, $i$, we compute a likelihood of the form:

$$ L_i = \prod_{j=1}^{N_{\gamma}^{obs}} [{G(\theta_j^{obs}| \theta_i^{exp})} + B]. $$

\noindent Here, $G$ is a Gaussian-like function which expresses the probability of observing the $j^{th}$ photon at an angle 
$\theta_j^{obs}$ with respect to the expected Cerenkov angle $\theta_i^{exp}$ for particle type $i$. The second term, $B$, is 
the background probability distribution which is a flat function in $\theta_i^{exp}-\theta_j^{obs}$. 
Using the likelihoods, we compute differences in negative-log likelihoods between two particle-type hypotheses: proton-kaon, proton-pion, and kaon-pion (i.e., ${\mathcal L}(K-\pi) = -2\log (L_K/L_{\pi})$). Defined in this manner, true kaons will peak
at negative values of ${\mathcal L}$ and true pions will peak at positive values of ${\mathcal L}$.

   Proton candidates are identified by requiring either $|S_{p}|<3$, $|S_{\pi}|>3$ and $|S_K|>3$, or at least three reconstructed
Cerenkov photons with ${\mathcal L}(p-\pi)<-5$ and ${\mathcal L}(p-K)<0$.  
Kaon candidates are only required to have either $|S_K|<3$ and $|S_{\pi}|>3$, or
at least three reconstructed Cerenkov photons with ${\mathcal L}(K-\pi)<0$.
Because pions are the majority particle in hadronic events, only loose
particle identification requirements are imposed. Pion candidates are required to have $|S_{\pi}|<4$ or at least three reconstructed Cerenkov photons with ${\mathcal L}(K-\pi)>-5$ and ${\mathcal L}(p-\pi)>-5$. By construction, the list of pion, kaon and proton candidates are not exclusive of each other, but in forming a given $\Lambda_c^+$ candidate, all daughter tracks must be distinct from one another. 

\begin{figure}[bht]
\vspace{0.0cm}
%\centerline{\epsfig{figure= lam0_pi0_prl.eps,height=8.0in}}
\centerline{\epsfig{figure=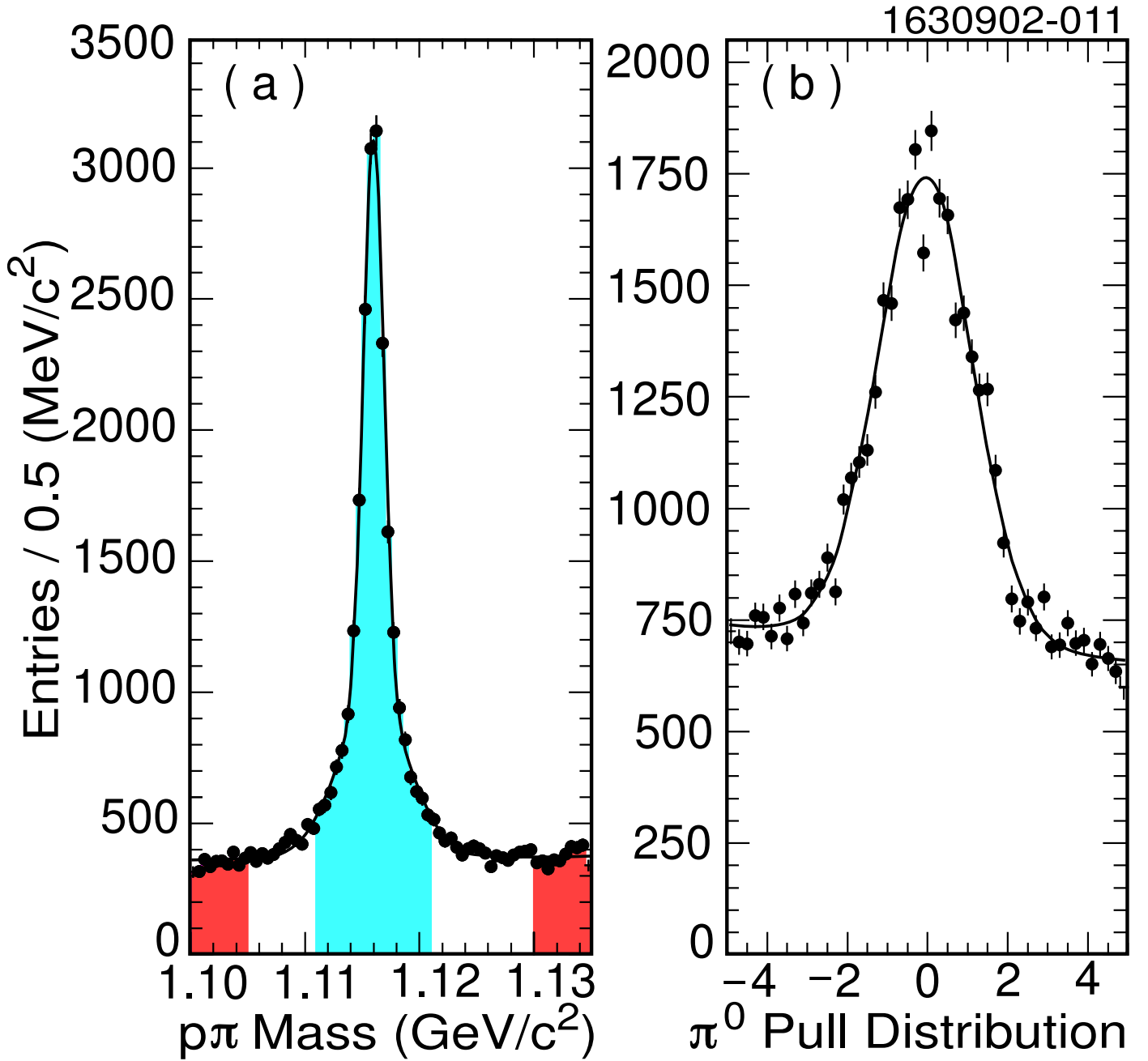,height=7.0in}}
%\vspace{0.0cm}
\caption{\label{lam0_pi0_prl} (a) Proton-pion invariant mass distribution after the proton identification requirement. The lightly-shaded area corresponds to the $\Lambda$ signal region and the darker shaded areas are the $\Lambda$ sidebands. (b) The pull distribution for $\pi^0$ mesons which have momentum larger than 250 MeV/$c$ and are in our $\Lambda_c^+$ candidate sample. The superimposed curves represent fits to the distributions as described in the text.}
\end{figure}

	 Candidate $\Lambda$ baryons are constructed by pairing oppositely charged proton and pion candidates and performing a constrained vertex fit. To suppress combinatorial background we require $\Lambda$ candidates to have a decay length of at least 1 cm. The invariant mass distribution of $\Lambda$ candidates is shown in Fig.~\ref{lam0_pi0_prl}(a). The distribution is fit using two Gaussians (with the same mean) to describe the signal and a first order polynomial to describe the background. The relative fraction of events in the Gaussian components is consistent with simulation. The signal (sideband) region includes candidates which have a mass in the range from 1111-1121 MeV/$c^2$ (1100-1105 or 1130-1135 MeV/$c^2$). The $\Lambda$ sidebands are used to check for a possible selection bias. Candidate $\pi^0$ mesons are formed by combining pairs of photons and requiring that the pull is less than 2. The pull is defined as the difference in the $\gamma\gamma$ invariant mass from the known $\pi^0$ mass, divided by the uncertainty in the $\gamma\gamma$ mass, where the uncertainty on the invariant mass is computed on a combination-by-combination basis. The pull distribution for the candidate $\pi^0$'s in the $\lamctolam4pi$ sample are shown in Fig.~\ref{lam0_pi0_prl}(b). Superimposed is the result of a fit to the sum of a Gaussian and a linear background. The fit has a mean of zero and a width of 1.1 units which is consistent with simulation. To reduce combinatorial background, we 
require that the energy of each photon is larger than 50 MeV and that the normalized energy asymmetry 
of the two photons $|E_1-E_2|/(E_1+E_2)<0.8$.

    Candidates for the decay $\lamctolam4pi$ are formed by combining $\Lambda$ candidates with one neutral and three charged pions. The combinatoric background is suppressed by requiring $\Lambda_c^+$ candidates to have a total momentum in excess of 3.5 GeV/$c$. Kinematically, this removes the contribution of $\Lambda_c^+$ baryons from $B$ meson decay, leaving only the contribution from charm fragmentation. To improve the signal-to-noise ratio,
we require $\Lambda$, $\pi^0$ and $\pi^{\pm}$ candidates to have momenta in excess of
500 MeV/c, 250 MeV/c, and 100 MeV/c respectively.

   Candidates for the $\lamctopkpi$ decay (reference mode) are formed by combining proton candidates with charged kaon and charged pion candidates. As with the $\lamctolam4pi$ candidates, we require that the $\lamctopkpi$ candidates have total momentum larger than 3.5 GeV/$c$.
The charge of the kaon is required to be opposite to that of the proton. As the decay length of the $\Lambda_c^+$ is too small to be observed in the CLEO detector, we require that all three tracks are consistent with coming from the $e^+e^-$ interaction point. We also require that the proton have a total momentum larger than 500 MeV/$c$, while the kaon and pion are required to have momenta larger than 300 MeV/$c$.

\begin{figure}[bht]
\vspace{0.0cm}
%\centerline{\epsfig{figure=lamc_lam4pi_data_prl.eps,height=8.0in}}
\centerline{\epsfig{figure=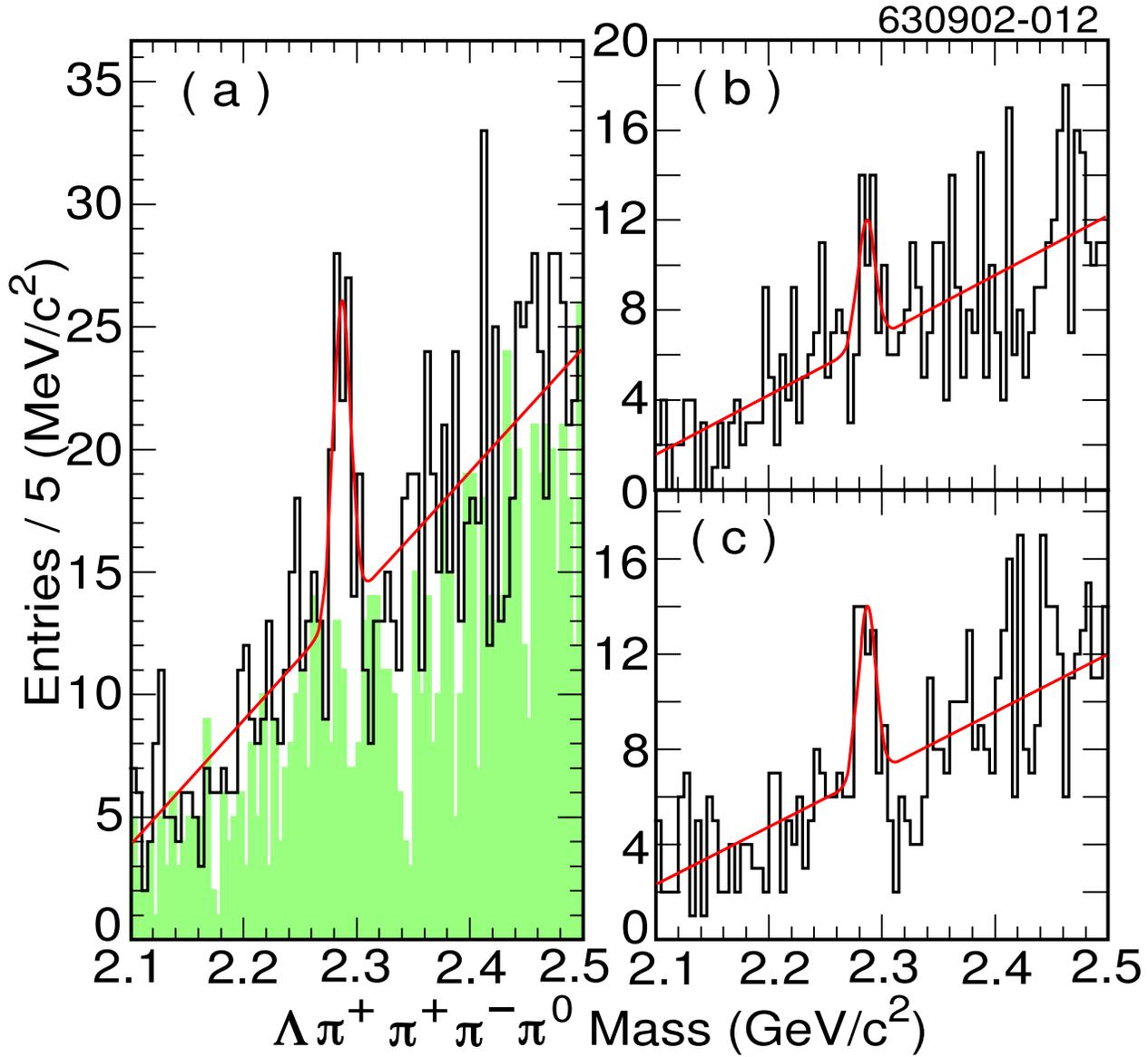,height=7.0in}}
%\vspace{0.0cm}
\caption{\label{lamc_lam4pi_data_prl} Invariant mass distributions for
(a) both $\Lambda\pi^+\pi^+\pi^-\pi^0$ and $\overline{\Lambda}\pi^-\pi^-\pi^+\pi^0$ combinations, (b) $\overline{\Lambda}\pi^-\pi^-\pi^+\pi^0$ combinations only, and (c) $\Lambda\pi^+\pi^+\pi^-\pi^0$ combinations only. In (a), the shaded distribution corresponds to candidates formed from the improper decay sequences ($\Lambda_c^+\to\overline{\Lambda}\pi^+\pi^+\pi^-\pi^0$ and $\overline{\Lambda_c^+}\to\Lambda\pi^-\pi^-\pi^+\pi^0$). The superimposed curves are fits to a Gaussian signal function and a linear background as described in the text.}
\end{figure}

	Figure~\ref{lamc_lam4pi_data_prl}(a) shows the invariant mass distribution for $\lamctolam4pi$ candidates for data after all selection criteria.  A clear excess of events is observed at the $\Lambda_c^+$ mass. The distribution is fit to the sum of a Gaussian whose width is fixed to the expected width from simulation (8.2 MeV/$c^2$) and a linear background from which we extract a signal yield of 49.5$\pm$11.9 events at a mass of 2287$\pm$1.3 (stat.) MeV/$c^2$. Also shown in the figure (shaded histogram) are combinations where we pair the $\pi^+\pi^+\pi^-\pi^0$ with a $\overline{\Lambda}$. No $\Lambda_c^+$ signal should be present in this mode, and the data show no significant excess in this mass range. The data have also been separated into $\Lambda_c^+$ and $\overline{\Lambda_c^+}$ in Fig.~\ref{lamc_lam4pi_data_prl}(b) and (c), respectively. They are fit using the same functional form as for the entire signal. The extracted yields are 21.3$\pm$8.1 $\Lambda_c^+$ and 28.3$\pm$8.5 $\overline{\Lambda_c^+}$ events. We therefore confirm that $\Lambda_c^+$ and $\overline{\Lambda_c^+}$ appear in approximately equal numbers. 

\begin{figure}[bht]
\vspace{0.0cm}
%\centerline{\epsfig{figure=omega_eta_invmass_prl.eps,height=8.0in}}
\centerline{\epsfig{figure=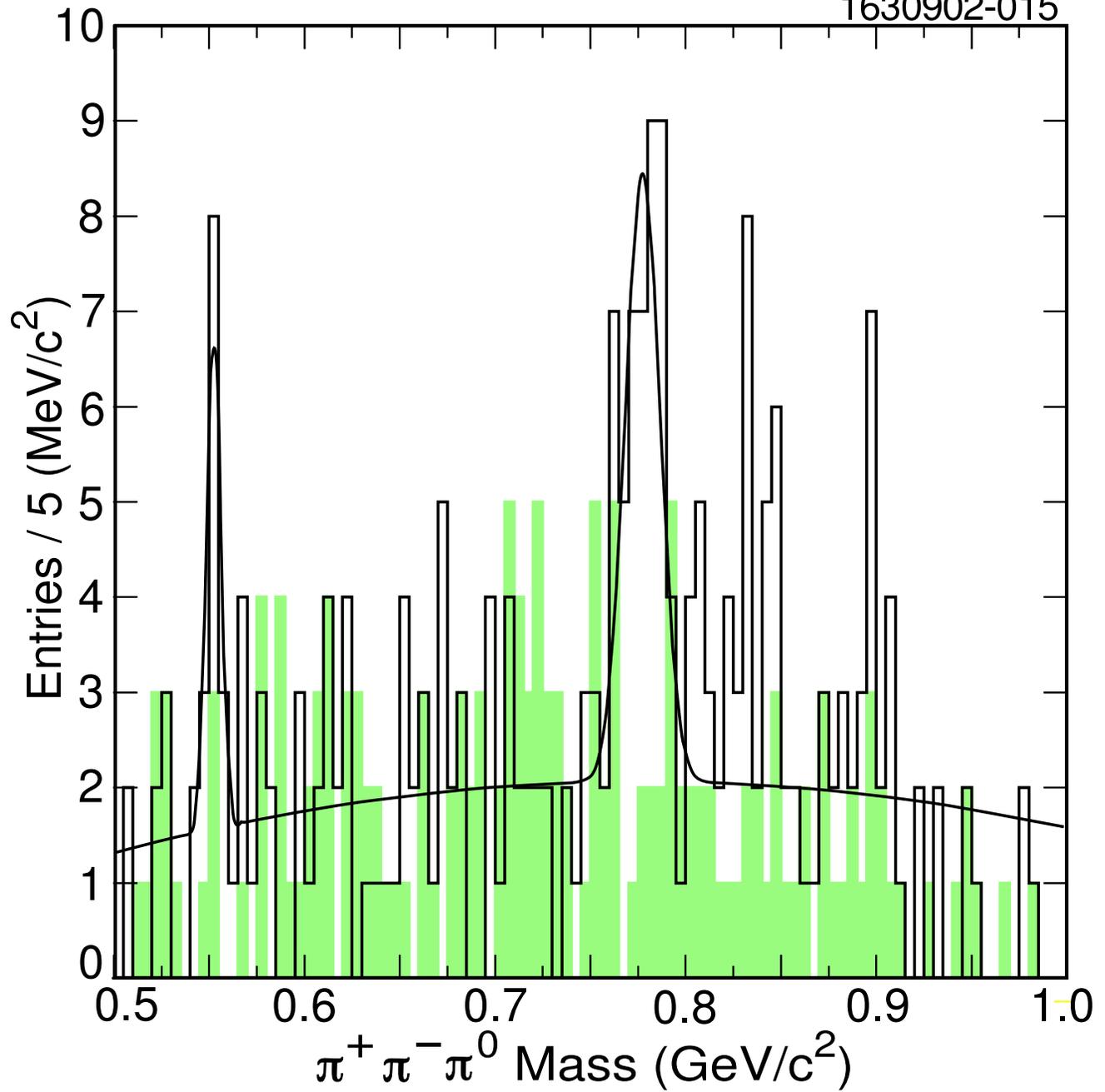,height=7.0in}}
%\vspace{0.0cm}
\caption{\label{omega_eta_invmass_prl} Invariant mass distribution for the two $\pi^+\pi^-\pi^0$ combinations in the $\lamctolam4pi$ decay for the $\Lambda_c^+$ signal region (histogram, 2270-2300 MeV/$c^2$) and the $\Lambda_c^+$ sidebands (shaded, 2245-2260 and 2320-2335 MeV/$c^2$). The superimposed curve represents a fit to the distribution as described in the text.}
\end{figure}

	We search for substructure in the $\Lambda\pi^+\pi^+\pi^-\pi^0$ final state by forming $\pi^+\pi^-\pi^0$ combinations (2 per event) among the 4 pions using $\Lambda\pi^+\pi^+\pi^-\pi^0$ combinations which have an invariant mass between 2270 and 2300 MeV/$c^2$. The $\pi^+\pi^-\pi^0$ invariant mass distributions are shown in Fig.~\ref{omega_eta_invmass_prl} for the $\Lambda_c^+$ signal region (2270-2300 MeV/$c^2$) and the $\Lambda_c^+$ sideband regions (2245-2260 and 2320-2335 MeV/$c^2$). A strong peak is observed near $M_{\omega}=782$ MeV/$c^2$ and a smaller excess at $M_{\eta}=547$ MeV/$c^2$.
The distribution is fit to the sum of two Gaussians, one for the $\omega$ and one for the $\eta$, and a second-order polynomial for the background. The signal shapes used in the fit are determined from simulation. The fitted masses are 552$\pm$2 MeV/$c^2$ and 778$\pm$3 MeV/$c^2$, consistent with the known masses of the $\eta$ and $\omega$ mesons. The probability that the enhancement at the $\eta$ mass is caused by a background fluctuation is about 4\%. The $\omega$ signal is approximately 6 standard deviations above the expected fitted background. We also observe three narrow peaks between 825 and 900 MeV/$c^2$. Each of these peaks correspond to $\sim$1.5-2 standard deviations above the background prediction, depending on the assumed Gaussian width. Since there are no known particles at these masses, we assume they are statistical fluctuations.
 
\begin{figure}[bht]
\vspace{0.0cm}
%\centerline{\epsfig{figure=lamc_omega_eta_prl.eps,height=8.0in}}
\centerline{\epsfig{figure=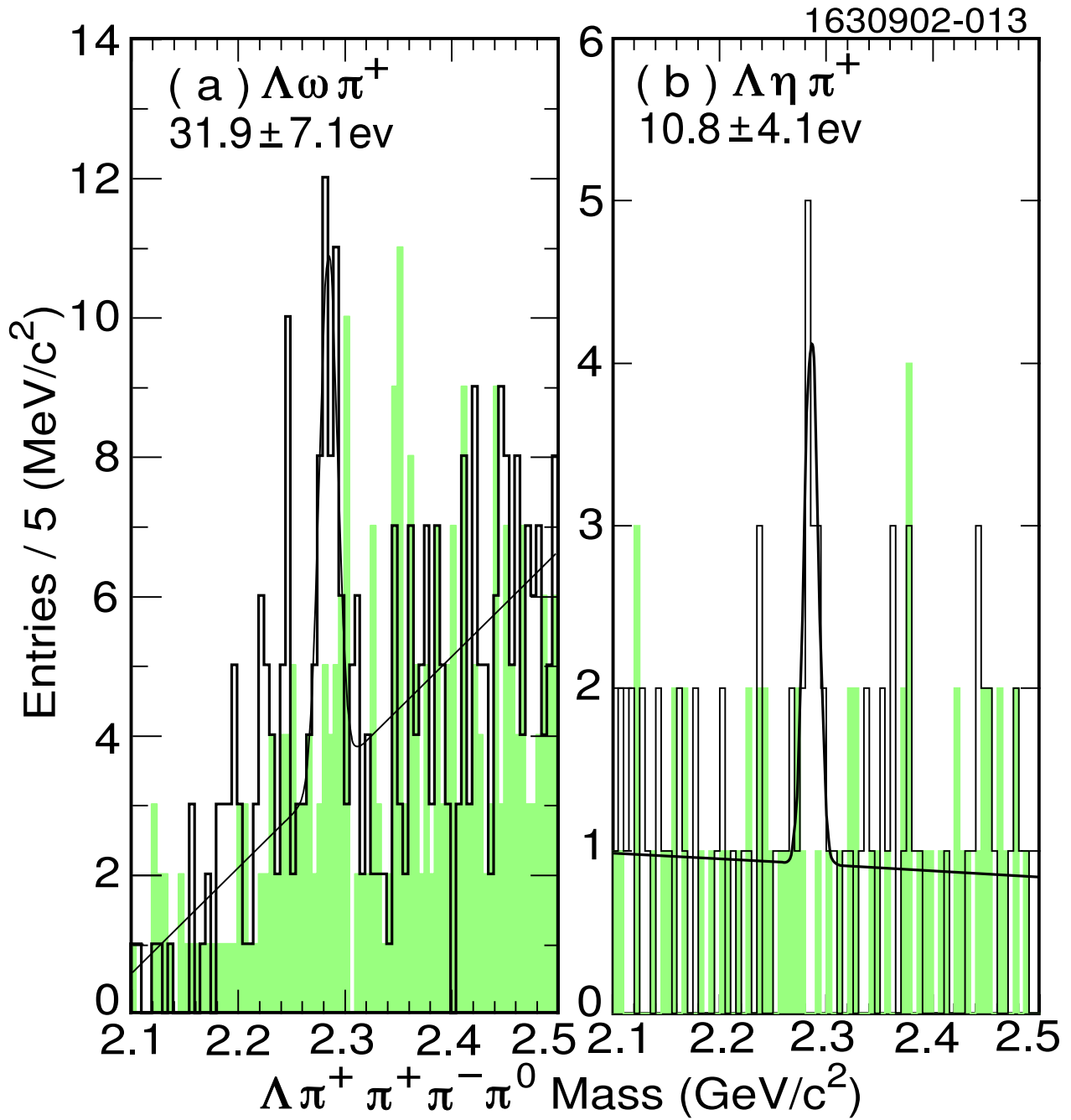,height=7.0in}}
%\vspace{0.0cm}
\caption{\label{lamc_omega_eta_prl} Invariant mass distribution for $\Lambda\pi^+\pi^+\pi^-\pi^0$ combinations from data where either of the two $\pi^+\pi^-\pi^0$ combinations have (a) a mass in the $\omega$ mass region (752 to 812 MeV/$c^2$) or (b) in the $\eta$ mass region (534 to 560 MeV/$c^2$). The shaded regions show the corresponding distributions for the $\omega$ sidebands (690-720 MeV/$c^2$ or 843-873 MeV/$c^2$) and the $\eta$ sidebands (500-513 MeV/$c^2$ or 580-593 MeV/$c^2$). The superimposed curves are fits to a Gaussian signal function and a linear background as described in the text. The fitted numbers of
events are shown at the top of each figure.}
\end{figure}

	The signal yields in the $\omega$ ($\eta$) modes are determined by requiring the $\pi^+\pi^-\pi^0$ invariant mass (either combination) to fall within $\pm$30 MeV/$c^2$ ($\pm$14 MeV/$c^2$) of their known masses and fitting for a signal in the $\Lambda\pi^+\pi^+\pi^-\pi^0$ invariant mass spectrum. To ensure non-overlapping sub-samples, we exclude $\lamctolametapi$ candidates from the $\lamctolamompi$ sample. The signal regions for the $\omega$ and $\eta$ correspond to about $\pm$3.5 times the Gaussian width obtained from simulations of these decay modes. The $\Lambda\pi^+\pi^+\pi^-\pi^0$ invariant mass spectra for the $\omega$ and $\eta$ modes are shown in Fig.~\ref{lamc_omega_eta_prl}(a) and (b), respectively. The shaded regions show the corresponding distributions when we select combinations which fall into the $\omega$ sidebands (690-720 MeV/$c^2$ or 843-873 MeV/$c^2$) and $\eta$ sidebands (500-513 MeV/$c^2$ or 580-593 MeV/$c^2$) respectively. The distributions are fit to the sum of a Gaussian and a linear background. The widths of the Gaussians are constrained to 8.4 MeV/$c^2$ for the $\omega$ and 6.6 MeV/$c^2$ for the $\eta$. These widths are determined 
from Monte Carlo studies of these processes, which includes the Breit-Wigner width of the $\omega$ and detector resolution effects. The fitted yields are 31.9$\pm$7.1 $\lamctolamompi$ candidates and 10.8$\pm$4.1 $\lamctolametapi$ candidates. Due to the limited statistics in the $\lamctolametapi$ mode, we also quote an
upper limit of 16.8 events at the 90\% confidence level. Thus these two modes account for 87\% of the
$\lamctolam4pi$ candidates. 

\begin{figure}[bht]
\vspace{0.0cm}
%\centerline{\epsfig{figure=lamc_omega_plus_eta_prl.eps,height=8.0in}}
\centerline{\epsfig{figure=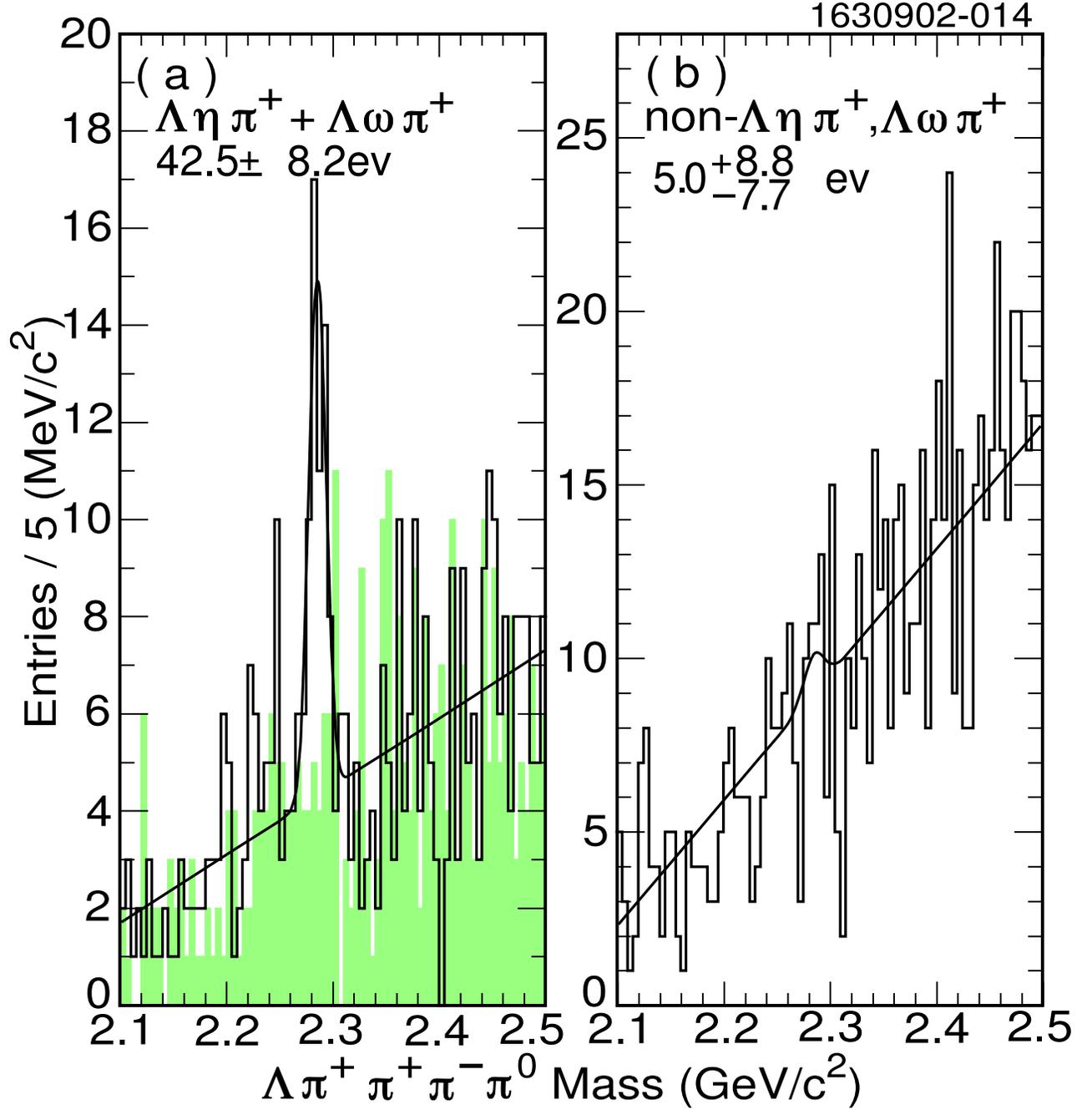,height=7.0in}}
%\vspace{0.0cm}
\caption{\label{lamc_omega_plus_eta_prl} Invariant mass distribution for $\Lambda\pi^+\pi^+\pi^-\pi^0$ combinations from data where either of the two $\pi^+\pi^-\pi^0$ combinations have (a) a mass in the $\omega$ or $\eta$ mass regions, and (b) all combinations for which the $\pi^+\pi^-\pi^0$ invariant mass falls outside the $\omega$ and $\eta$ mass regions. The shaded region in (a) shows the corresponding distributions for the $\omega$ and $\eta$ sidebands (see text for signal and sideband region definitions). The superimposed curves are fits to a Gaussian signal function and a linear
background as described in the text. The fitted numbers of
events are shown at the top of each figure.}
\end{figure}

	The contribution from the $\omega$ and $\eta$ are combined and shown in Fig.~\ref{lamc_omega_plus_eta_prl}(a). The spectrum is fit to the sum of a Gaussian and a linear background shape. The width is constrained to the event-weighted average of the widths obtained from the $\lamctolamompi$ and $\lamctolametapi$ simulations (7.9 MeV/$c^2$). The number of fitted candidates is 42.5$\pm$8.2, which agrees with the sum from the individual fits. Events which do not have a $\pi^+\pi^-\pi^0$ with invariant mass in the $\eta$ or $\omega$ region are shown in Fig.~\ref{lamc_omega_plus_eta_prl}(b). We fit the spectrum assuming a Gaussian signal shape and a first order polynomial for the background shape. The mean of the Gaussian is fixed to the world-average value (2285 MeV/$c^2$) and the width is constrained to the value found from a simulation of non-resonant $\lamctolam4pi$ (9.8 MeV/$c^2$). The fitted yield of $5.0^{+8.4}_{-7.7}$ events is consistent with the expected difference of 7.0 events as well as zero. This corresponds to an upper limit of 17.7 events at the 90\% confidence level.

	The fitted $\lamctolamompi$ and $\lamctolametapi$ signals are corrected for possible contributions from non-resonant production by subtracting off the expected non-resonant contribution, as determined from simulation. The number of non-resonant events is taken to be 7.0 events as discussed above. This number must be corrected for the contribution from $\lamctolamompi$, because some of the signal lies outside the $\omega$ mass window. The  $\eta$ width is negligible compared to the mass resolution, and therefore no such correction is necessary for $\lamctolametapi$. From simulation, we find that (3$\pm$2)\% of the $\omega$ signal events fall outside the $\omega$ mass window. We therefore estimate that 6.0 of the 49.5 events are from sources other than $\lamctolamompi$ and $\lamctolametapi$. We assume that these 6 events are from non-resonant sources, and using simulation we find that 16\% (4\%) are expected to have a $\pi^+\pi^-\pi^0$ combination in the $\omega$ ($\eta$) mass window. The corrected yields are therefore 31.0$\pm$7.1 and 10.6$\pm$4.1 for the $\lamctolamompi$ and $\lamctolametapi$ modes. The 3\% loss of $\omega$ events due to the mass window definition are accounted for in the $\lamctolamompi$ efficiency. We take the systematic uncertainty on the non-resonant subtraction to be 100\% of the correction.

	We have investigated the possibility that the $\omega\pi$ system may come from higher mass resonance decays, such as $\Lambda_c^+\to\Lambda\rho^{\prime +}, \rho^{\prime +}\to\omega\pi^+$, which may contribute to this final state due to its large width. We compared the $\omega\pi^+$ invariant mass distribution in data to a non-resonant $\omega\pi^+$ simulation and the expectation from $\rho^{\prime +}\to\omega\pi^+$. No peaks are observed in data, and a Kolmogorov-Smirnov test of the shapes give a consistency of 80\% with respect to a pure non-resonant $\Lambda\omega\pi^+$ final state and a 10\% consistency with pure $\Lambda\rho^{\prime +}$. Thus, the data are more compatible with a pure non-resonant $\omega\pi^+$ contribution, although a $\rho^{\prime +}$ contribution cannot be ruled out. 

	The relative branching fraction ${\cal B}( \lamctolam4pi )/{\cal B}( \lamctopkpi)$ is:

\vspace{-0.5cm}
\begin{center}
$$ \frac{{\cal B}( \lamctolam4pi )}{{\cal B}( \lamctopkpi)} =    
{N( \lamctolam4pi ) \over 
N( \lamctopkpi ) } 
{\epsilon (\lamctopkpi ) \over
\epsilon ( \lamctolam4pi ) },$$
\end{center}

\noindent
where $N( \lamctolam4pi )$ and $ N( \lamctopkpi )$ are the number of reconstructed $\lamctolam4pi$ and $\lamctopkpi$ signal events and $\epsilon( \lamctolam4pi )$ and $\epsilon( \lamctopkpi )$ are the relative efficiencies for reconstructing these modes respectively.
	
	The efficiencies for the $\Lambda\pi^+\pi^+\pi^-\pi^0$ and the $pK\pi$ final states are obtained using a Monte Carlo simulation of these decays 
(see Table~\ref{table:1}). Events are simulated using the Lund/Jetset model~\cite{lund} for the production mechanism ($e^+ e^- \to q\overline{q}$) and are allowed to decay according to the known lifetime and decay branching fractions~\cite{PDG2000}. The response of the detector to the particles in the generated events is simulated using a GEANT-based simulation~\cite{geant} of the detector. The fully simulated events are then passed through the same analysis chain as data. For the $pK\pi$ final state, our simulation includes the resonant sub-modes $pK^*$ and $\Delta^{++} K$ as well as a non-resonant $pK\pi$ contribution~\cite{PDG2000}. The efficiency for the $\lamctolam4pi$ final state is taken to be the branching fraction weighted efficiency for $\lamctolamompi$, $\lamctolametapi$, and non-resonant $\lamctolam4pi$. Table~\ref{table:1} summarizes the inputs for the computation of the branching fractions. 
	
	The branching fractions for $\lamctolam4pi$, $\lamctolamompi$ and $\lamctolametapi$ relative to $\lamctopkpi$ are measured to be:

\begin{center}
$$ \frac{{\cal B}( \lamctolam4pi )}{{\cal B}( \lamctopkpi )} = 0.36 \pm 0.09\pm 0.09$$
$$\frac{{\cal B}( \lamctolamompi , \omega\to\pi^+\pi^-\pi^0 )}{{\cal B}( \lamctopkpi )}=0.21\pm 0.05\pm0.05$$
$$\frac{{\cal B}( \lamctolametapi, \eta\to\pi^+\pi^-\pi^0  )}{{\cal B}( \lamctopkpi )}=0.093\pm 0.038\pm0.022 
{\hspace{0.1in}\rm( <0.15~at~90\% CL)},$$
\end{center}

\noindent where the first uncertainty is statistical and the second is systematic. The systematic uncertainties are discussed below.

%Using ${\cal B}(\eta,\omega\to \pi^+\pi^-\pi^0)$  from Table~\ref{table:1} we obtain: 
%
%\vspace{-0.5cm}
%\begin{center}
%$$ \frac{{\cal B}( \lamctolamompi )}{{\cal B}( \lamctopkpi )}=0.24\pm 0.06 $$
%$$ \frac{{\cal B}( \lamctolametapi )}{{\cal B}( \lamctopkpi )}=0.41\pm 0.17$$
%\end{center}

	These measurements are the first observations of $\lamctolam4pi$ modes, which are (or almost) entirely through $\lamctolamompi$ and $\lamctolametapi$. This is the first observation of the $\lamctolamompi$ decay mode. The $\lamctolametapi$ decay has been previously observed in $\eta\to\gamma\gamma$ decays~\cite{lametapi}, but this is the first indication in the $\eta\to\pi^+\pi^-\pi^0$ mode. Our measurement compares well with those results which yielded ${\cal B}( \lamctolametapi )/ {\cal B}( \lamctopkpi )=0.35\pm 0.05$(stat.)$\pm$0.06(syst.).

	As a systematic check of our procedure, we have also measured the relative branching fraction for $\lamctolampi$ and find ${\cal B}( \lamctolampi )/ {\cal B}( \lamctopkpi )$=0.17$\pm$0.02(stat.). This is consistent with the current world average of 0.18$\pm$0.03~\cite{PDG2000}. 

	Significant systematic uncertainties arise from sources which affect the $\Lambda\pi^+\pi^+\pi^-\pi^0$ mode but not the $p K^- \pi^+$ mode (or vice versa). Systematic uncertainties include estimation of reconstruction efficiencies of tracks, $\Lambda$ and $\pi^0$ reconstruction, particle identification selection, background estimation and signal estimation.

	Studies of data and simulation show that the charged particle tracking simulation reproduces data within 2\%. A 2\% systematic uncertainty per track introduces a 6\% relative systematic uncertainty. The systematic uncertainty in
the $\pi^0$ reconstruction efficiency is estimated
by comparing the differential yield for $\Upsilon(4S)\to\pi^0 X$ in CLEO III data to CLEO II data. The difference is found to be 8\% for our selection criteria. When combined with the 5\% systematic uncertainty on the CLEO II $\pi^0$ efficiency, we assign a 10\% uncertainty to the $\pi^0$ reconstruction efficiency. The systematic uncertainty in the $\Lambda$ reconstruction efficiency is estimated by comparing our measured value for ${\cal B}(\lamctolampi )/{\cal B}(\lamctopkpi)$ 
[0.17$\pm$0.02(stat.)] to the world average value (0.18$\pm$0.03). While these results are consistent
with one another, we take the uncertainty to be the the quadrature sum of the difference in the
two results (6\%) and the relative uncertainty in the world average value (17\%). The uncertainty
in the  $\Lambda$ reconstruction efficiency is therefore estimated to be 18\%. The systematic uncertainty in the particle identification was investigated using a sample of protons from $\Lambda\to p\pi^-$ and kaons from $D^{*+}\to D^0( \to K^-\pi^+ )\pi^+$. The studies indicate that the systematic uncertainty is less than 3\% for protons and less than 1\% for kaons. The uncertainty in the background is estimated by using alternate background shapes derived from (a) the $\overline{\Lambda}\pi^+\pi^+\pi^-\pi^0$ invariant mass distribution in data and (b) the $\Lambda\pi^+\pi^+\pi^-\pi^0$ invariant mass distribution using the $\Lambda$ sidebands. We also varied the number of bins used in the fit to the background; these sources lead to an uncertainty of 6\% on the signal yield. 

Additional uncertainty in the signal estimation may arise from an imperfect understanding of the signal resolution and efficiency. The resolution uncertainty is estimated by varying the widths by one standard deviation ($\approx \pm$ 1 MeV/$c^2$) and refitting the data. The uncertainty in the signal yield is found to be 5\%. For the efficiency uncertainty, we consider one source which only affects the full $\lamctolam4pi$ branching fraction measurement, and one which affects the resonant $\omega$ and $\eta$ contributions. For the full $\lamctolam4pi$ sample, we use a branching fraction weighted efficiency. To evaluate the systematic uncertainty, we shift the number of $\omega$ events by $\pm$7.6 events and the corresponding number of $\eta$ events by $\mp$7.6 events and compute the corresponding changes in the $\lamctolam4pi$ branching fraction. We find a shift of $\pm$3\% in the $\lamctolam4pi$ branching fraction. For the resonant modes, we also include a 3\% systematic which corresponds to a 100\% relative uncertainty on the subtraction for possible non-resonant contributions which have a $\pi^+\pi^-\pi^0$ invariant mass in either the $\omega$ or $\eta$ mass region. A compilation of the sources of systematic uncertainty is presented in Table~\ref{systematics}. The total systematic uncertainty 
is estimated to be 13\%. 

	These measurements indicate that ($87\pm 13$)\% of the four-pion state is feed-down through 
resonant $\omega$ and $\eta$ production. The branching fractions (after dividing through 
by the ${\cal B}(\omega\to\pi^+\pi^-\pi^0)$ and ${\cal B}(\eta\to\pi^+\pi^-\pi^0)$) are shown in
Table~\ref{table:3}.  The common systematic uncertainty in the branching fraction for $\lamctopkpi$ 
has not been included in the last column in Table~\ref{table:3}. The $\lamctolametapi$ mode has been 
measured previously in the $\eta\to\gamma\gamma$ mode~\cite{lametapi}, and the result presented 
here is consistent with those findings. We find that the branching fraction for $\lamctolamompi$ is
(0.24$\pm$0.06$\pm$0.06) of ${\cal B}( \lamctopkpi )$. For the $\eta$ mode, we find a rate relative 
to ${\cal B}( \lamctopkpi )$ 
of (0.41$\pm$0.17$\pm$0.10) with an upper limit of 0.65 at 90\% CL. The non-$\omega,\eta$ signal 
constitutes $5.0^{+8.4}_{-7.7}$ events which corresponds to an upper limit of 17.7 events at 
90\% confidence level. This corresponds to an upper limit on the branching fraction relative to 
${\cal B}(\lamctopkpi$) of 0.13 at 90\% confidence level.

	The fact that the $\Lambda\pi^+\pi^+\pi^-\pi^0$ rates are competitive with $\Lambda\pi^+$ ((0.9$\pm$0.3)\%), $\Lambda\pi^+\pi^0$ ((3.6$\pm$1.3)\%) and $\Lambda\pi^+\pi^-\pi^+$ ((3.3$\pm$1.0)\%) is not surprising, as charm mesons exhibit similar branching fractions for $D\to K n( \pi)$, where $n=1-4$. In fact, it is interesting to note that the exclusive rate for $D^0\to K^-\omega\pi$ ((2.7$\pm$0.5)\%) constitutes about 68\% of the rate for $D^0\to K^-\pi^+\pi^+\pi^-\pi^0$ ((4.0$\pm$0.4)\%). This is quite comparable to what we find here for the ratio of the exclusive $\Lambda\omega\pi^+$, $\omega\to\pi^+\pi^-\pi^0$ rate to the total $\Lambda\pi^+\pi^+\pi^-\pi^0$ rate (59\%).

	We gratefully acknowledge the effort of the CESR staff in providing us with
excellent luminosity and running conditions.
M. Selen thanks the PFF program of the NSF and the Research Corporation, 
and A.H. Mahmood thanks the Texas Advanced Research Program.
This work was supported by the National Science Foundation, and the
U.S. Department of Energy.

\newpage

\begin{table}[htb] 
%\begin{center}  
\caption{Inputs for the computation of the branching fractions for $\lamctolam4pi$, $\lamctolamompi$ and $\lamctolametapi$. 
The label ``tot'' refers to the total $\lamctolam4pi$ sample for which the
efficiency is computed as a weighted average efficiency of efficiencies as described 
in the text. The label ``nr'' refers to the simulated non-resonant $\lamctolam4pi$ sample.}
\label{table:1} 
\begin{tabular}{@{}cccc}
\hline 
Mode & Data Yield   & Eff. (\%) & ${\cal B}$ ($\pi^+\pi^-\pi^0$)(\%) \\ 
\hline
$p K \pi$    		&   1959$\pm$57   & 20.8$\pm$0.9  &    --   \\
$\Lambda\pi^+\pi^+\pi^-\pi^0$ (tot) & 49.5$\pm$11.9 &  1.46$\pm$0.10 &   --   \\ 
$\Lambda\omega\pi^+$  & 31.0$\pm$7.6 & 1.56$\pm$0.10 & 88.8$\pm$0.7  \\
$\Lambda\eta\pi^+$    & 10.6$\pm$4.1 & 1.22$\pm$0.10 & 22.6$\pm$0.4  \\
$\Lambda\pi^+\pi^+\pi^-\pi^0$ (nr) & 7.9 &  1.48$\pm$0.10 &   --   \\ 
\hline 
\end{tabular} 
%\end{center} 
\end{table}

\begin{table}[htb] 
%\vspace{0.25in}
%\begin{center}  
\caption{Systematic uncertainties in the measurement of the branching fractions for $\lamctolam4pi$, $\lamctolamompi$ and $\lamctolametapi$.} \label{systematics} 
\begin{tabular}{lc}\hline 
Source & Value \\ \hline 
Background shape    &   $\pm$6\%  \\
Signal Width	  &   $\pm$6\%  \\
Track Recon. Efficiency & $\pm$6\% \\
$\pi^0$ Recon. Efficiency & $\pm$10\% \\
Proton ID efficiency & $\pm$3\% \\
Kaon ID efficiency & $\pm$1\% \\
$\Lambda$ Recon. Efficiency & $\pm$18\% \\
Non-resonant subtraction($\omega$, $\eta$ only) & $\pm$3\% \\
Signal Efficiency($\lamctolam4pi$ only) & $\pm$3\% \\
\hline
Total  &	$\pm$24\% \\
\hline 
\end{tabular} 
%\end{center} 
\end{table} 

\begin{table}[htb]
 
%\begin{center}  
\caption{Summary of measured $\Lambda_c^+$ branching fractions. The uncertainties shown are statistical (first)
and systematic (second). The qualifier ``tot'' refers to the total yield in the $\Lambda\pi^+\pi^+\pi^-\pi^0$ 
final state and ``nr'' refers to non-$\omega$ and non-$\eta$ contributions. For the $\lamctolametapi$ mode, we show
both the central value and the 90\% confidence upper limit (in parentheses).} 
\label{table:3} 
\begin{tabular}{@{}ccccc}
\hline 
Mode & & ${\cal B}({\rm Mode})/{\cal B}( \lamctopkpi )$   & & ${\cal B}$(Mode) (\%) \\
\hline
$\lamctolam4pi$ (tot) & & $0.36\pm 0.09\pm 0.09$ & & $1.79\pm 0.47\pm 0.43$ \\
$\lamctolamompi$  & & $0.24\pm 0.06\pm 0.06$ & & $1.19\pm 0.30\pm 0.29$ \\
$\lamctolametapi$ & & $0.41\pm 0.17\pm 0.10$ ($<$ 0.65) & & $2.06\pm 0.84\pm 0.49$ ($<$ 3.3 ) \\
$\lamctolam4pi$ (nr) & & $<0.13$ & & $<0.65$ \\
\hline 
\end{tabular} 
%\end{center} 
\end{table}

\newpage

\end{document}